# Polarization Angle Scanning for Wide-band Millimeter-wave Direct Detection

Heyao Wang, Ziran Zhao, Lingbo Qiao, and Dalu Guo

*Abstract*—Millimeter-wave (MMW) technology has been widely utilized in human security screening applications due to its superior penetration capabilities through clothing and safety for human exposure. However, existing methods largely rely on fixed polarization modes, neglecting the potential insights from variations in target echoes with respect to incident polarization. This study provides a theoretical analysis of the cross-polarization echo power as a function of the incident polarization angle under linear polarization conditions. Additionally, based on the transmission characteristics of multi-layer medium, we extended the depth spectrum model employed in direct detection to accommodate scenarios involving multi-layered structures. Building on this foundation, by obtaining multiple depth spectrums through polarization angle scanning, we propose the Polarization Angle-Depth Matrix to characterize target across both the polarization angle and depth dimensions in direct detection. Simulations and experimental validations confirm its accuracy and practical value in detecting concealed weapons in human security screening scenarios.

*Index Terms*—Cross-polarization, polarization angle scanning, concealed weapon, millimeter-wave, direct detection

## I. INTRODUCTION

Millimeter-wave (MMW) technology offers superior penetration through clothing while ensuring safe radiation levels, making it an excellent choice for concealed weapon detection. In recent years, MMW technology has been extensively studied and applied in human security screening [1]. The polarization characteristics of electromagnetic waves contain valuable information about the structure and material properties of targets. Multi-polarization detection, which introduces additional dimensions of information [2][3], has become increasingly central to improving detection accuracy. Furthermore, under a single polarization condition, different polarization directions result in variations in the information acquired [4][5], making multi-polarization detection an increasingly important research focus.

In passive MMW detection, two-dimensional Stokes vector images obtained through fully polarized radar have been employed for analyzing target clustering characteristics [6].

This work was supported in part by Beijing Nova Program *(Corresponding author: Ziran Zhao.)*

Heyao Wang, Ziran Zhao, Lingbo Qiao and Dalu Guo are with the Department of Engineering Physics, Tsinghua University, Beijing 100084, China, and also with the National Engineering Research Center of Dangerous Articles and Explosives Detection Technologies, Beijing 100084, China (e-mail: zhaozr@tsinghua.edu.cn).

Image quality can be improved by averaging images acquired from three different polarization directions [7]. Additionally, by collecting images under different linear polarization directions and calculating specific parameters, previous studies have explored various applications, including imaging quality enhancement [8], target contour extraction [9], and contrast improvement [10][11].

In active MMW detection, dual-polarization imaging systems have been integrated into holographic imaging systems for human security screening, effectively improving detection accuracy [12]. Since concealed objects exhibit significantly different polarization characteristics compared to the human body, their co-polarization and cross-polarization echoes vary accordingly. By extracting co-polarization and cross-polarization echo information and computing polarization-related parameters such as the polarization ratio, concealed objects can be identified [13]-[17]. Furthermore, by performing H-α decomposition on the full-polarization information of a target, material classification can be achieved based on its location in the H-α plane [18]. Several studies have further explored the differences between human body and weapons in co-polarization and cross-polarization images by comparing four-channel images under horizontal and vertical polarization [19]-[22].

However, most of the mentioned studies rely on fixed polarization modes (e.g., horizontal, vertical, or ±45°) and have not thoroughly investigated the variations in target echoes with respect to the incident polarization angle. To address this limitation, this study analyzes the variation in cross-polarization echo power as a function of the incident polarization angle and derives its theoretical expression. The depth spectrum utilized in millimeter-wave (MMW) direct detection is based on wide-band power information and represents the distribution of the target along the depth direction [16]. Based on the transmission characteristics of multi-layer medium, we extended the depth spectrum model to better align with real-world scenarios. Building upon this, we examine the differences in cross-polarization responses and depth distribution between human surfaces and concealed weapons. By scanning the incident polarization angle and obtaining multiple depth spectrums, we introduce the Polarization Angle-Depth Matrix to characterize the features in polarization response and depth distribution. Through both simulations and experiments, the proposed analysis is validated, demonstrating its potential application value in human security screening for concealed weapon detection.

The remainder of this paper is structured as follows: Section

II investigates the variation in cross-polarization echo power with incident polarization angle and explores the depth-wise distribution characteristics of multi-layer targets, providing corresponding theoretical formulations. The differences between human bodies and weapons in these two dimensions are further analyzed, and the Polarization Angle-Depth Matrix is introduced for feature representation. Section III presents simulation and experimental results, verifying the theoretical analysis and demonstrating the potential application value of the proposed matrix in human security screening. Section IV concludes the paper with a summary of key findings.

## II. Theory and Characteristic Analysis of Polarization Angle-Depth Matrix

### A. Variation of Cross-Polarization Power with Incident Linear Polarization Angle

Previous study [23] has shown that the human body surface can be approximately modeled as a planar or cylindrical surface. When linearly polarized waves are incident on the human body surface, the scattered echoes predominantly exhibit co-polarized components, while cross-polarized components are generally weak. However, for weapons, due to their relatively complex structures, the cross-polarized components in the scattered echoes are significantly stronger. Based on this observation, individuals carrying concealed weapons exhibit notable differences in cross-polarized echo power compared to ordinary individuals. Meanwhile, due to the differences in the polarization scattering matrices of weapons and the human body surface, the variation of cross-polarization echo power with incident polarization angle also differs. This section will theoretically derive this variation pattern.

In the horizontal-vertical coordinate system $h-v$, the polarization scattering matrix of a target at a given incident angle is defined as $\boldsymbol{S} = \begin{bmatrix} s_{hh} & s_{hv} \\ s_{vh} & s_{vv} \end{bmatrix}$, where the incident wave is expressed as $\boldsymbol{E}_T = \begin{bmatrix} E_{Th} \\ E_{Tv} \end{bmatrix} = \begin{bmatrix} \cos\eta \\ \sin\eta \end{bmatrix}$, here $\eta$ is the polarization angle. The scattered echo can then be written as

$$\boldsymbol{E}_R = \boldsymbol{S}\boldsymbol{E}_T = \begin{bmatrix} s_{hh}\cos\eta + s_{hv}\sin\eta \\ s_{vh}\cos\eta + s_{vv}\sin\eta \end{bmatrix} = \begin{bmatrix} E_{Rh} \\ E_{Rv} \end{bmatrix} \quad (1)$$

To analyze the cross-polarization, a new orthogonal polarization basis, denoted as $h'-v'$, is introduced. Here, $h'$ represents the polarization direction of the incident wave. This basis is equivalent to rotating the original $h-v$ coordinate system by an angle $\eta$ about the origin. By transforming $\boldsymbol{E}_R$ using the rotation matrix $\boldsymbol{R} = \begin{bmatrix} \cos\eta & \sin\eta \\ -\sin\eta & \cos\eta \end{bmatrix}$, the cross-polarization component can be obtained as $E'_{Rv} = E_{Rv}\cos\eta - E_{Rh}\sin\eta$. The cross-polarization power is then defined as

$$\begin{aligned} P_{cross}(\eta) &= \|E'_{Rv}\|^2 \\ &= \|E_{Rh}\|^2 \sin^2\eta + \|E_{Rv}\|^2 \cos^2\eta \\ &\quad -(E^*_{Rv}E_{Rh} + E^*_{Rh}E_{Rv})\sin\eta\cos\eta \end{aligned} \quad (2)$$

Define coefficients as

$$\begin{aligned} a_1 &= |s_{hv}|^2 \\ a_2 &= 2Re\{s^*_{hv}(s_{hh} - s_{vv})\} \\ a_3 &= |s_{hh}|^2 + |s_{vv}|^2 - 2(Re\{s^*_{hv}s_{vh}\} + Re\{s^*_{hh}s_{vv}\}) \\ a_4 &= 2Re\{s^*_{vh}(s_{vv} - s_{hh})\} \\ a_5 &= |s_{vh}|^2 \end{aligned} \quad (3)$$

then we obtain

$$\begin{aligned} P_{cross}(\eta) &= \frac{1}{8}(3a_1 + 3a_5 + a_3) \\ &\quad + \frac{1}{4}[2(a_5 - a_1)\cos 2\eta + (a_2 + a_4)\sin 2\eta] \\ &\quad + \frac{1}{8}[(a_1 + a_5 - a_3)\cos 4\eta + (a_4 - a_2)\sin 4\eta] \end{aligned} \quad (4)$$

Define another group of coefficients as

$$\begin{aligned} A_1 &= \frac{1}{4}\sqrt{4(a_5 - a_1)^2 + (a_2 + a_4)^2} \\ B_1 &= -\arctan\left(\frac{a_2 + a_4}{2(a_5 - a_1)}\right) \\ A_2 &= \frac{1}{8}\sqrt{(a_1 + a_5 - a_3)^2 + (a_4 - a_2)^2} \\ B_2 &= -\arctan\left(\frac{a_4 - a_2}{a_1 + a_5 - a_3}\right) \\ C &= \frac{1}{8}(3a_1 + 3a_5 + a_3) \end{aligned} \quad (5)$$

and we can obtain a further simplified form

$$P_{cross}(\eta) = A_1\cos(2\eta + B_1) + A_2\cos(4\eta + B_2) + C \quad (6)$$

According to (6), the following properties can be derived:
1) The variation of the target's cross-polarization power with incident polarization angle is expressed as the sum of two cosine functions with periodicities of π and π/2, respectively. The amplitudes and phase shifts of these cosine functions are determined by the polarization scattering matrix of the target.
2) If the target is rotated by an angle $\theta$ around the incident direction (i.e., the z-axis), it is equivalent to keeping the target fixed while rotating the incident wave polarization angle by $-\theta$. This results in a translation of (6), leading to the following equivalent expression:

$$\begin{aligned} P_{cross}(\eta) &= A_1\cos(2(\eta + \theta) + B_1) \\ &\quad + A_2\cos(4(\eta + \theta) + B_2) + C \end{aligned} \quad (7)$$

3) For most common weapons, such as metallic handguns, the co-polarization components of the polarization scattering matrix are typically 1–2 orders of magnitude higher than the cross-polarization components, i.e., $s_{hh}, s_{vv} \gg s_{hv}, s_{vh}$. Under this condition, the amplitude coefficients in (6) satisfy $A_2 \gg A_1$. Consequently, the $A_2 \cos(4\eta + B_2)$ term in (6) becomes the dominant component, while the $A_1 \cos(2\eta + B_1)$ term introduces only minor amplitude and phase modulations.

*B. Depth Spectrum Theory of multi-layer medium*

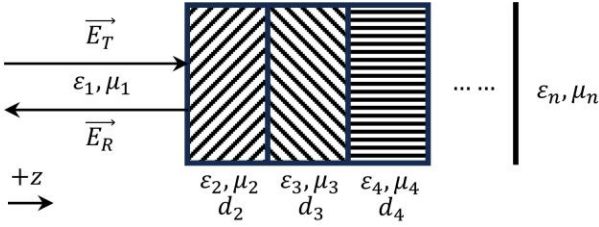

**Fig. 1.** Plane wave incidence on a multi-layered medium.

For scenarios where dangerous items are carried on the human body, the situation can be equivalently modeled as a multi-layered medium placed in front of a semi-infinite medium (the human body), as illustrated in Fig. 1. Stuart W. Harmer et al. provided a theoretical expression for a multi-scatter point model, which bears similarity to the multi-layered medium. By utilizing only the echo power information, they derived a mathematical expression that characterizes the separations between the scatter points, referred to as the depth spectrum [16]:

$$s'(\tau) = \sum_{u=1}^{n-1} \sum_{i=1}^{n-1} F^{-1}\{a_i(f)a_u^*(f)\} \otimes \delta\left(\tau - \frac{2}{c}(z_i - z_u)\right)$$
$$= \sum_{i=1}^{n-1} F^{-1}\{|a_i(f)|^2\}\delta(\tau)$$
$$+ \sum_{u=2}^{n-1} \sum_{i=1}^{n-u} F^{-1}\{a_i^*(f)a_{i+u-1}(f)\}\delta\left(\tau - \frac{2}{c}(z_{i+u-1} - z_{i+1})\right) \quad (8)$$

where $a_i$ represents the scattering strength, $z_i$ denotes the distance between scatter point $i$ and the transceiver antenna, $F^{-1}$ represents the inverse Fourier transform and $\otimes$ represents the convolution operator.

However, (8) cannot intuitively represent the characteristics of the depth spectrum in a multi-layer medium model. This limitation arises because (8) is based on the assumption that scattering points are independent of each other. In multi-layer medium, the intensity of the incident wave at each interface is influenced by the transmission and attenuation effects of the preceding layers. As a result, the scattering coefficient $a_i$ at one specific interface inherently contains an exponential attenuation term related to the material properties and thickness of all preceding layers. After performing the inverse Fourier transform, this attenuation term also exhibits the form of a δ-function, which merges with the original terms in (8). Consequently, in multi-layer medium, the positions of the δ-functions in the depth spectrum no longer directly correspond to the actual physical distances between interfaces but rather to the optical depth.

S. W. Harmer et al. theoretically derived the depth spectrum form for a single wax layer placed on the surface of the human body in [15]. This paper further derives the depth spectrum form for multi-layer medium.

Assuming that a uniform plane wave is normally incident on the surface of a multi-layered isotropic medium. Under the assumption of single reflection at each interface, the equivalent reflection coefficient of the entire medium structure can be expressed as

$$R = Q_1 + \sum_{i=2}^{n-1} Q_i \exp\left(-j2 \sum_{q=2}^{i} \beta_q d_q\right) \quad (9)$$

where $Q_i = \prod_{p=1}^{i-1} r_{i\,i+1} t_{p\,p+1} t_{p+1\,p} \exp\left(-2\sum_{q=2}^{i} \alpha_q d_q\right)$ for $i \geq 2$, and $Q_1 = r_{12}$. Here $r_{i\,i+1}$ represents the reflection coefficient of the incident wave in the +z-direction at the interface between medium $i$ and $i+1$. Similarly, $t_{i\,i+1}$ represents the transmission coefficient of the incident wave in the +z-direction at the same interface. The complex wavenumber in medium $i$ is defined as $\kappa_i = \frac{2\pi f}{c}\sqrt{\varepsilon_i \mu_i} = \beta_i + j\alpha_i$, where $\alpha_i$ is the attenuation factor, and $\beta_i$ is the phase factor.

Assuming that within the frequency band of interest $\varepsilon_i$ is constant, $\alpha_i$ is negligible (i.e., $\varepsilon_i \approx Re\{\varepsilon_i\}$), and $\mu_i = 1$, the phase factor becomes $\beta_i = \frac{2\pi f}{c}\sqrt{\varepsilon_i}$. Under these conditions, the reflection coefficient simplifies to $r_{i\,i+1} = \frac{\sqrt{\varepsilon_i} - \sqrt{\varepsilon_{i+1}}}{\sqrt{\varepsilon_i} + \sqrt{\varepsilon_{i+1}}}$, which is also constant.

In a monostatic transceiver configuration, the frequency-domain representation of the target's scattered echo can be expressed as $S(f) = A(f)R(f)\exp\left(-j\frac{4\pi f}{c}z_0\right)$, where $A(f)$ is the gain coefficient determined by the radiation patterns of the transmitting and receiving antennas. This can be normalized using strong scatterers such as metallic plates. Here, $z_0$ denotes the distance between the target and the transceiver system. The normalized time-domain signal can be expressed as $s_n(t) = F^{-1}[S(f)] = F^{-1}[R(f)] \otimes \delta\left(t - \frac{2z_0}{c}\right)$.

Using (9), we can derive

$$F^{-1}[R(f)] = Q_1 + \sum_{i=2}^{n-1} Q_i \delta\left(t - \frac{2\sum_{q=2}^{i}\sqrt{\varepsilon_q}d_q}{c}\right) \quad (10)$$

Substituting (10) into the time-domain signal expression, we obtain

$$s_n(t) = Q_1 \delta\left(t - \frac{2z_0}{c}\right)$$
$$+ \sum_{i=2}^{n-1} Q_i \delta\left(t - \frac{2(z_0 + \sum_{q=2}^{i}\sqrt{\varepsilon_q}d_q)}{c}\right) \quad (11)$$

At the receiver, the measured broadband power signal is $|S(f)|^2$. Performing the inverse Fourier transform on it is equivalent to computing the autocorrelation of the time-domain echo signal (ignoring constant scaling factors of the

transform):

$$s'(\tau) = F^{-1}[|S(f)|^2] = s_n^*(-t) \otimes s_n(t) \quad (12)$$

By combining (11) and (12), we arrive at (negative time-axis terms are discarded)

$$s'(\tau) = \sum_{i=1}^{n-1} |Q_i|^2 \delta(\tau)$$
$$+ \sum_{u=2}^{n-1} \sum_{i=1}^{n-u} Q_i^* Q_{i+u-1} \delta\left(\tau - \frac{2\sum_{q=i+1}^{i+u-1} \sqrt{\varepsilon_q} d_q}{c}\right) \quad (13)$$

Compared to (8), the δ-function positions in the depth spectrum of the multi-layer medium, as expressed in (13), correspond to $\sqrt{\varepsilon_q} d_q$, which is referred to as the optical thickness, and their amplitudes are determined by the reflection and transmission coefficients of each layer.

*C. Characteristics of the Polarization Angle-Depth Matrix*

Based on the previous analysis, to characterize the differences between the human body surface and weapons in both polarization response and depth dimensions, we propose the Polarization Angle-Depth Matrix as a data representation. Since the cross-polarization reflection and transmission coefficients of the target are dependent on the polarization direction of the incident wave, we denote the incident wave polarization angle as $\eta$. Consequently, the depth spectrum (13) should be rewritten as:

$$s'_X(\tau, \eta) = \sum_{i=1}^{n-1} |Q_{Xi}(\eta)|^2 \delta(\tau)$$
$$+ \sum_{u=2}^{n-1} \sum_{i=1}^{n-u} Q_{Xi}^*(\eta) Q_{X\,i+u-1}(\eta) \delta\left(\tau - \frac{2\sum_{v=i+1}^{i+u-1} \sqrt{\varepsilon_v} d_v}{c}\right) \quad (14)$$

Here, the subscript **X** denotes cross-polarization. In practical applications, by discretizing $\eta$ and $\tau$, (14) can be expressed in matrix form as:

$$\boldsymbol{U}_X = \begin{bmatrix} s'_X(\tau_1, \eta_1) & \cdots & s'_X(\tau_M, \eta_1) \\ \vdots & \ddots & \vdots \\ s'_X(\tau_1, \eta_N) & \cdots & s'_X(\tau_M, \eta_N) \end{bmatrix} \quad (15)$$

Equation (15) is referred to as the Polarization Angle-Depth Matrix of the target. Further analysis will be conducted on the two-dimensional characteristics of (15).

1) The column dimension of $\boldsymbol{U}_X$ represents the incident polarization angle. The first column can be expressed as:

$$s'_X(0, \eta) = \sum_{i=1}^{n-1} |Q_{Xi}(\eta)|^2 \delta(0) \quad (16)$$

According to the analysis in Section A, (16) follows the variation pattern described in (6) with respect to the incident polarization angle. For different targets, the characteristics of their polarization scattering matrices result in significant differences in the amplitude and phase of the cosine functions.

For the human body surface, since $s_{hv}$ and $s_{vh}$ are approximately zero, the cross-polarization reflection coefficient is very small, leading to minimal variation in $s'_X(0, \eta)$ with respect to η. This implies that the amplitude terms $A_1$ and $A_2$ in (6) are small. Furthermore, for a standing human body, the rotational angle around the incident direction remains fixed, leading to a constant phase in (6), i.e., $B_1$ and $B_2$ remain nearly unchanged.

For weapons, the values of $s_{hv}$ and $s_{vh}$ increase significantly, leading to a much higher overall cross-polarization power compared to the human body surface. As a result, the amplitude of the cosine functions in (6) increases notably. Additionally, due to variations in the rotational angle of the weapon around the incident direction, the phase terms in (6) may differ from those of the human body.

2) The rows of $\boldsymbol{U}_X$ represent the depth dimension. Denoting the depth spectrum at a specific incident polarization angle $\eta_0$, it can be rewritten as

$$s'_X(\tau, \eta_0) = \sum_{i=1}^{n-1} |Q_{Xi}(\eta_0)|^2 \delta(\tau)$$
$$+ \sum_{u=2}^{n-1} \sum_{i=1}^{n-u} Q_{Xi}^*(\eta_0) Q_{X\,i+u-1}(\eta_0) \delta\left(\tau - \frac{2\sum_{v=i+1}^{i+u-1} \sqrt{\varepsilon_v} d_v}{c}\right) \quad (17)$$

For the human body surface, since there is no multi-layer structure, the Dirac delta sequence in (17) contains only the first term, meaning that there is a single peak at $\tau = 0$.

For weapons, which can be equivalently modeled as a multi-layer structure, multiple peaks corresponding to the second term in (17) will appear in addition to the first peak. Furthermore, due to the increased cross-polarization reflection coefficient, the peak at $\tau = 0$ is significantly higher than that of the human body surface.

Based on the above analysis, the Polarization Angle-Depth Matrix exhibits distinct characteristics along both its column and row dimensions: The human body surface shows a smooth polarization response with a single peak in the depth spectrum. Weapons, on the other hand, exhibit a higher polarization response amplitude and contain multiple peaks in the depth spectrum. These characteristics provide a reliable basis for the effective detection of concealed weapons.

## III. RESULTS AND DISCUSSION

*A. Simulation Results*

To validate the effectiveness of the proposed analysis, we conducted a simulation using FEKO software. First, we verify the correctness of (5) and (6). We utilized a target composed of a combination of a rectangular dihedral reflector and a flat plate, as shown in Fig. 2. During the simulation, the dihedral reflector was rotated by certain angles around both the *z*-axis and the *y*-axis. A uniform plane wave was used as the incident wave to obtain the polarization scattering matrix **S** of the target. By scanning the incident polarization angle, we obtained multiple $(P_{cross}, \eta)$ data pairs. These data pairs were subsequently used for curve fitting. The curve fitting results were compared with the theoretical calculations based on the polarization scattering matrix, as shown in Fig. 3.

Fig. 3 demonstrates that the cross-polarization power curve obtained from theoretical calculations aligns closely with the results of cosine fitting based on the collected power data points. This confirms the validity of (5) and (6).

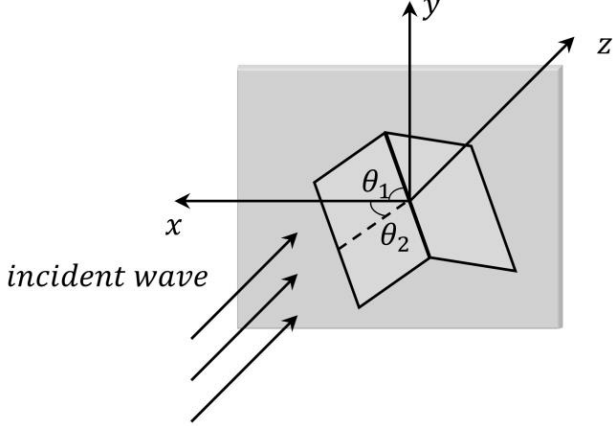

**Fig. 2.** Schematic of the simulation scenario (dihedral reflector and flat plate).

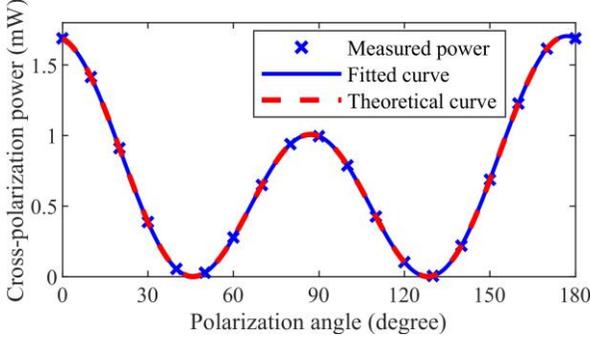

**Fig. 3.** Curve fitting and theoretical calculation results: $\theta_1 = -50°, \theta_2 = 20°$.

Next, we conducted a simulation analysis of the Polarization Angle-Depth Matrix. In FEKO software, we modeled two metallic grenades, a metallic handgun, and a human body, and performed simulations where the grenades and handgun were placed on the human body surface. These cases were compared with the scenario of a standalone human body, as illustrated in Fig. 4. In the simulation, the frequency range was set to 75–110 GHz, with a sampling interval of 1 GHz. The polarization angle range was set from 0° to 180°, with a sampling interval of 10°. The simulated Polarization Angle-Depth Matrix obtained from these scenarios are shown in Fig. 5.
.

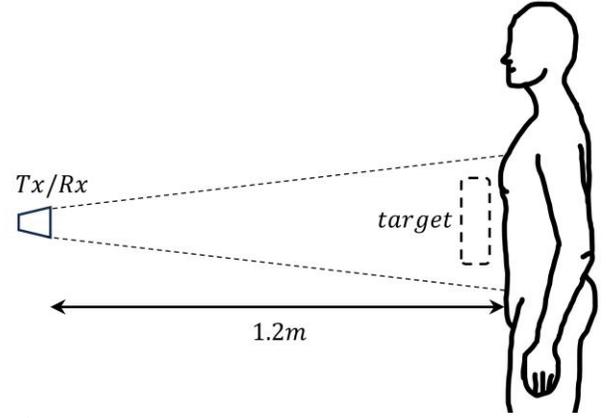

**Fig. 4.** Simulation scenario.

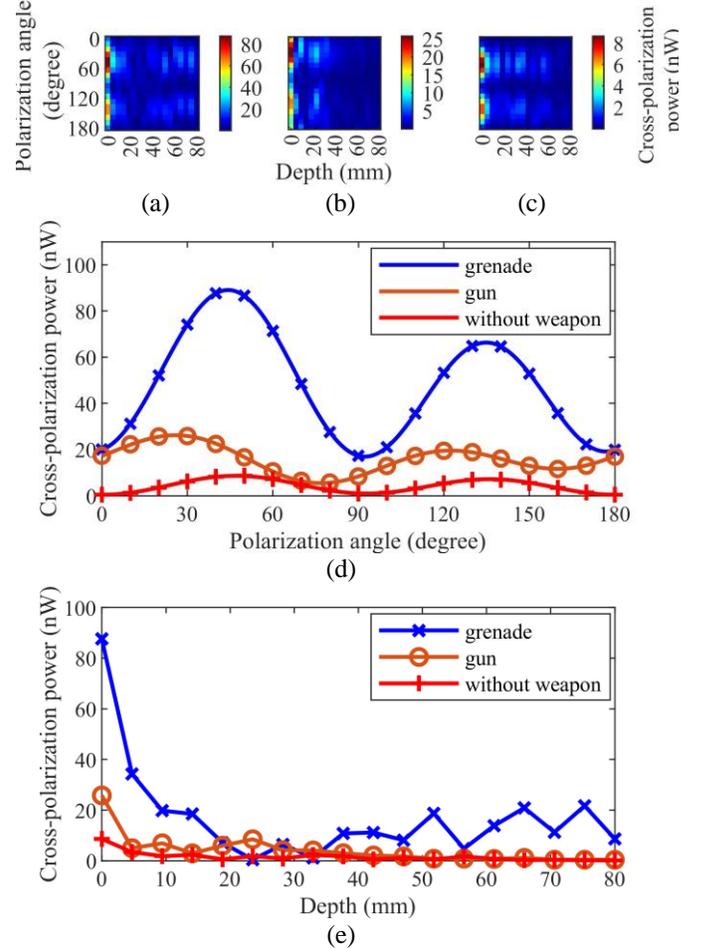

**Fig. 5** Simulation Results. Polarization Angle-Depth Matrix: (a) carrying two grenades; (b) carrying a handgun; (c) without weapons. (d) First column of the Polarization Angle-Depth Matrix. (e) Rows corresponding to the maximum value in the Polarization Angle-Depth Matrix of the three scenarios.

*B. Experiment Results*

We conducted experiments using a self-built experimental platform and a dummy model. The transmitting and receiving antennas were each fixed on two perpendicular rotary stages, allowing for the adjustment of the polarization angle via the rotation of the stages. In the experiments, grenade and metallic handgun models were selected as target objects. The measurement equipment included a vector network analyzer (VNA) and frequency extension modules, covering a frequency range of 75–110 GHz with a frequency step size of 0.175 GHz. The polarization angle range was set from 0° to 180°, with a sampling interval of 10°. The distance between the dummy and the antennas was 1.2 meters. The experimental platform as well as the weapons and clothing used in the experiment are shown in Fig. 6.

It is important to note that in the experiment, the transmitting and receiving antennas were placed side by side, rather than using a beam splitter to simulate a quasi-monostatic configuration. This was because the experiment only required power information and did not need phase information. By placing the two antennas in parallel, coupling between the transmitter and receiver was reduced, simplifying the experimental design and improving measurement stability.

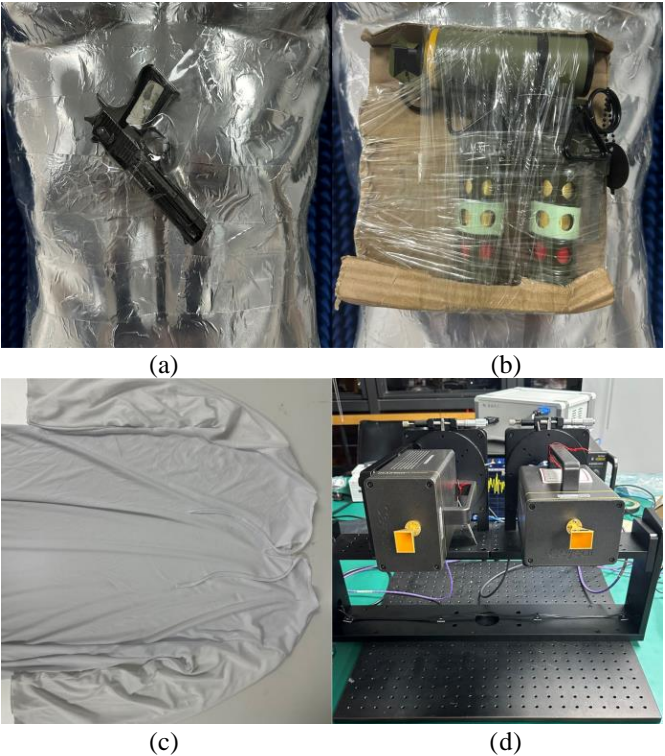

**Fig. 6.** Experiment scenario. (a) Handgun model. (b) Grenade model. (c) Clothing. (d) Experimental platform.

The Polarization Angle-Depth Matrix obtained from the experiment is shown in Fig. 7, where the cross-polarization echo power had been normalized according to the frequency response of the transmitted power.

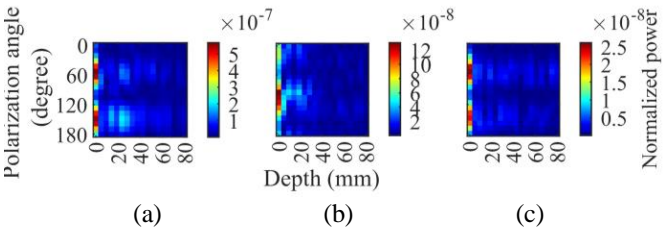

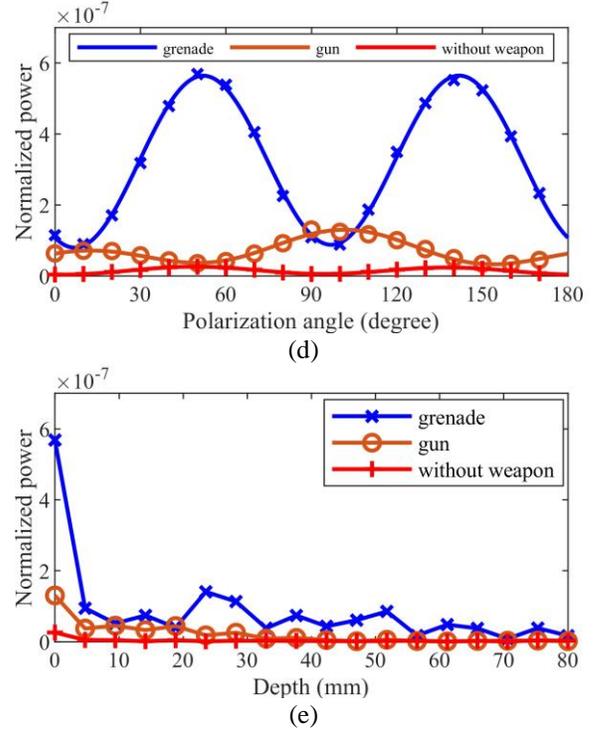

**Fig. 7** Experimental Results. Polarization Angle-Depth Matrix: (a) carrying grenade model; (b) carrying handgun model; (c) without weapons. (d) First column of the Polarization Angle-Depth Matrix. (e) Rows corresponding to the maximum value in the Polarization Angle-Depth Matrix of the three scenarios.

*C. Discussion*

The simulation results (Fig. 5) and experimental results (Fig. 7) demonstrate that individuals carrying different weapons exhibit significant differences in both dimensions of the Polarization Angle-Depth Matrix compared to those without weapons.

In the polarization angle dimension, the amplitude is significantly higher when carrying a weapon compared to the case without weapons. Additionally, the cosine phase differences among the three cases are also noticeable.

In the depth dimension, the peak values are significantly higher when carrying grenade model or handgun model compared to the case without weapons. Furthermore, the number of peaks increases notably, reflecting the influence of the weapon's structural characteristics on the depth distribution.

These results strongly indicate that the Polarization Angle-Depth Matrix can effectively characterize the differences in polarization response and depth distribution between individuals carrying weapons and normal individuals. This validates its potential application for concealed weapon detection.

VI. CONCLUSION

This study introduces the Polarization Angle-Depth Matrix by scanning incident polarization angle and capturing multiple depth spectrums in millimeter direct detection. Through the theoretical analysis of the variation of cross-polarization power with the incident polarization angle and the depth spectrum representation of multi-layer medium, it is observed that the human body surface and concealed weapons exhibit significant differences in polarization response and depth distribution. Simulation and experimental results validate that these features exhibit strong potential for detecting concealed weapons on individuals, highlighting its applicability in security screening.